\documentclass[sigconf,screen,authorversion]{acmart}

\setcopyright{rightsretained}
\copyrightyear{2024}
\acmYear{2024}
\acmDOI{10.1145/3639476.3639770}

\acmConference[ICSE-NIER'24]{New Ideas and Emerging Results}{April 14--20, 2024}{Lisbon, Portugal}
\acmBooktitle{New Ideas and Emerging Results (ICSE-NIER'24), April 14--20, 2024, Lisbon, Portugal}
\acmISBN{979-8-4007-0500-7/24/04}

\usepackage{enumitem}
\newcommand{\GraphGenForCode}{Graph\-Gen\-4\-Code} %

\begin{document}

\title{Towards Trustworthy AI Software Development Assistance}

\author{Daniel Maninger}
\orcid{0009-0005-0649-4958}
\affiliation{%
    \institution{Technische Universität Darmstadt}
    \department{Hessian Center for Artificial Intelligence (hessian.AI)}
    \city{Darmstadt}
    \country{Germany}}
\email{maninger@cs.tu-darmstadt.de}

\author{Krishna Narasimhan}
\orcid{0000-0001-8004-3470}
\affiliation{%
    \institution{AI Quality \& Testing Hub}
    \city{Frankfurt am Main}
    \country{Germany}}
\email{k.narasimhan@aiqualityhub.com}
\authornote{Work conducted while afiliated with Technische Universität Darmstadt.}

\author{Mira Mezini}
\orcid{0000-0001-6563-7537}
\affiliation{%
    \institution{Technische Universität Darmstadt}
    \department{Hessian Center for Artificial Intelligence (hessian.AI)}
    \department{National Research Center for Applied Cybersecurity ATHENE}
    \city{Darmstadt}
    \country{Germany}}
\email{mezini@cs.tu-darmstadt.de}

\renewcommand{\shortauthors}{Maninger et al.}

\begin{abstract}
    It is expected that in the near future, AI software development assistants will play an important role in the software industry. However, current software development assistants tend to be unreliable, often producing incorrect, unsafe, or low-quality code. We seek to resolve these issues by introducing a holistic architecture for constructing, training, and using trustworthy AI software development assistants. In the center of the architecture, there is a foundational LLM trained on datasets representative of real-world coding scenarios and complex software architectures, and fine-tuned on code quality criteria beyond correctness. The LLM will make use of graph-based code representations for advanced semantic comprehension. We envision a knowledge graph integrated into the system to provide up-to-date background knowledge and to enable the assistant to provide appropriate explanations. Finally, a modular framework for constrained decoding will ensure that certain guarantees (e.g., for correctness and security) hold for the generated code.
\end{abstract}



\received{14 September 2023}
\received[revised]{21 Dezember 2023}
\received[accepted]{22 November 2023}

\maketitle

\section{Introduction}\label{sec:intro}

Software development (SD) is a complex, expensive process~\cite{boehm2000software}. In times of ever-increasing digitalization of our society and scarcity of IT talent, it makes sense to (partially) automate this process -- the popularity that AI-assisted code generation has gained since the appearance of large language models (LLMs) indicates this need. At the same time, given that software dominates our digitized society, guaranteeing the quality of the generated code is paramount. Moreover, for the acceptance of AI software development assistants, it is essential that they cover a broad range of software engineering tasks and faithfully mimic existing well-understood and proven SD methods and practices.

Although existing LLMs, when pre-trained or fine-tuned on code\footnote{For brevity, we will use the term \emph{code model}.}, have shown promise in handling coding-related queries and synthesizing coherent-looking code, they exhibit concerning issues. For instance, AIs like ChatGPT produce erroneous code suggestions: On Stack Overflow, the large number of ChatGPT-generated inaccurate answers led to a ban on its use. The moderators stated that \textit{``the posting of answers created by ChatGPT [\dots] is substantially harmful to the site and to users who are asking questions and looking for correct answers''}\footnote{\raggedright\url{https://meta.stackoverflow.com/questions/421831/temporary-policy-chatgpt-is-banned}}. A later study made similar observations~\cite{kabir_who_2023}.

Moreover, a study on GitHub Copilot revealed that 40\% of the generated code contained critical security vulnerabilities~\cite{copilotstudy40}. Further emphasizing the dangers of such issues in the domain of software engineering is an investigation that discovered that \textit{``participants who had access to the AI assistant were more likely to introduce security vulnerabilities [\dots], yet also more likely to rate their insecure answers as secure''}~\cite{perry2022users}. These studies highlight quality and security problems with current industrial AI software development assistants.

To remedy these problems, we seek to establish the foundations of a next-generation AI software development assistant system\footnote{For brevity, we will also refer to it simply as \emph{(AI SD) assistant}.}. Given the software engineer's natural language queries, the assistant should suggest \emph{high-quality} code, i.e., correct but also secure, readable, and otherwise compliant with best practices. So far, code model research has focused almost exclusively on correctness (e.g.,~\cite{codebert, graphcodebert, travtrans, coderl, li2022competition, shiv_novel_2019}).

Beyond that, the assistant should be able to serve the role of a virtual pair programmer, explaining its suggestions, helping with debugging, etc., similar to how a human would. The positive effects of pair programming practices on code correctness and quality~\cite{pairprogramming1, pairprogramming2, pairprogramming3}, as well as on collaborative learning~\cite{nadi2016jumping}, are well known. Despite this, pair programming has not been widely adopted because working closely with a teammate is challenging\footnote{\url{https://martinfowler.com/articles/on-pair-programming.html\#Challenges}}. A virtual pair programmer would not cause such problems. Moreover, we envision the system we build as an AI counterpart to the traditional DevOps pipeline. Therefore, our assistant should be able to support the software engineer in all phases, from software design to deployment.

\begin{figure*}
    \centering
    \includegraphics[width=1.0\textwidth]{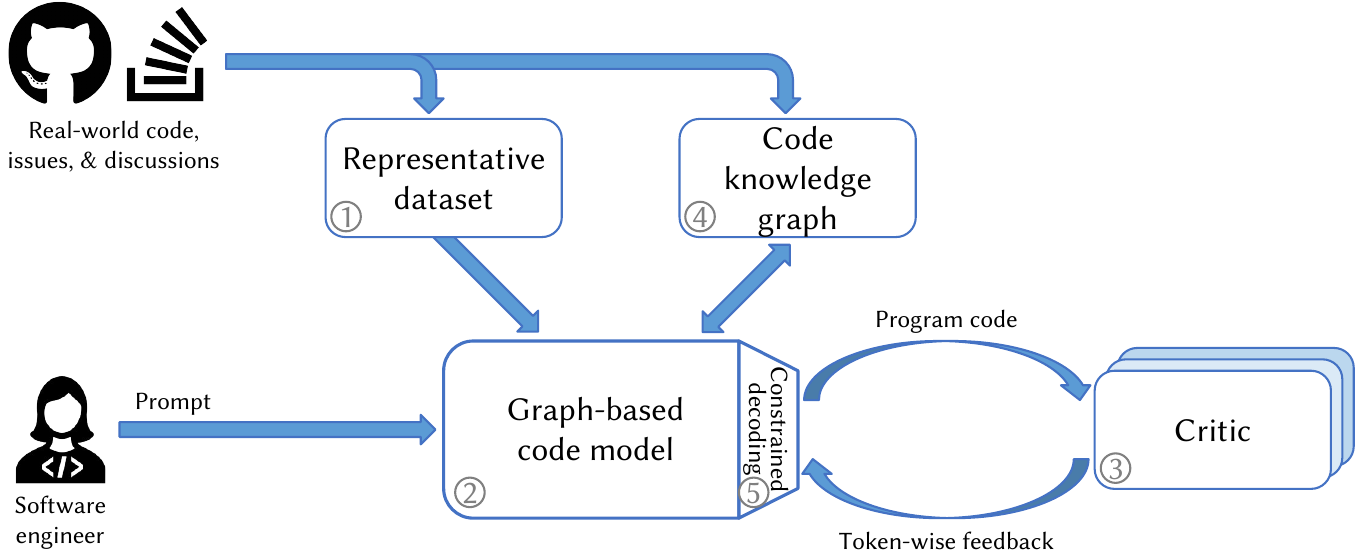} %
    \caption[Architecture]{High-level architecture of our envisioned AI software development assistant. It consists of five main components: (1)~A curated training \emph{dataset} representing real-world coding patterns and software architectures. (2)~A foundational \emph{code model} that uses \emph{graph representations} for better understanding of program semantics. (3)~An RL-based \emph{feedback mechanism} to fine-tune the model for improved code quality. (4)~A semantically enriched \emph{code knowledge graph} to help the assistant explain its code. (5)~A modular \emph{constrained decoding} framework on top of the model that prevents the generation of undesired code.}
    \label{fig:arch}
\end{figure*}

The listed objectives are highly ambitious and require addressing various challenges. This paper is intended to outline a way forward for us and other researchers trying to develop better AI software development assistants. To this end, we start by identifying five key \textbf{challenges}:
\begin{enumerate}[nosep]
    \item lack of representative datasets,
    \item difficulties capturing code structure and semantics,
    \item low code quality,
    \item insufficient explanations, and
    \item no guarantees on the results.
\end{enumerate}
We then propose five \textbf{solutions} to address these challenges:
\begin{enumerate}[nosep]
    \item curated real-world datasets,
    \item graph representations of code,
    \item fine-tuning through feedback from code analyses
    \item enriched code knowledge graphs, and
    \item constrained decoding.
\end{enumerate}
Our considerations lead to a holistic multi-component architecture for AI software development assistant systems built around a central code model. It is illustrated in Figure~\ref{fig:arch}.

\section{Approach}

The first step towards improving current AI software development assistants is to understand their limitations. Here, we elaborate on the five challenges we found to be most relevant. For each of them, we propose a possible solution together with an evaluation plan. All the described ideas contribute to the overall goal of this vision: Creating a \emph{trustworthy} AI SD assistant that a software engineer can use in real-world software projects without having to worry about the correctness, quality, safety, and security of the generated code.

\subsection{Representative Datasets}\label{subsec:dataset}

\paragraph{Challenge}

Popular code datasets\footnote{With closed-source commercial code models, the datasets used are often unknown.} such as CodeSearchNet\footnote{\url{https://github.com/github/CodeSearchNet}}~\cite{husain2019codesearchnet} and Py150K\footnote{\url{https://www.sri.inf.ethz.ch/py150}} consists of individual code snippets with single functions. Such datasets do not represent real-world software, typically exhibiting complex software patterns, interdependent class hierarchies, project dependencies involving multiple interconnected files, and further complexities. A study by Hellendoorn et~al.~\cite{whenfail} revealed that the absence of these features in coding benchmarks contributed to many shortcomings of code completion systems. We, therefore, consider the lack of high-quality datasets that accurately represent real-world coding patterns and complex software architectures a key challenge for training code models. However, real-world code suitable for creating such datasets is scarce. Moreover, annotating and curating code datasets requires significant code quality and dataset management expertise.

\paragraph{Envisioned solution}

To meet this challenge, we plan to compile and curate a comprehensive dataset of code that accurately reflects common real-world coding patterns and software structures. For this, we will build upon established state-of-the-art datasets.
To make these datasets representative, we will devise heuristics rooted in general coding patterns observed in real-world coding practices. Hellendoorn et~al. introduced initial heuristics for the C\# programming language to distinguish real-world completions from synthetic ones, such as length, type, and origin of (syntactic) tokens. We will expand on these heuristics in a generalized manner for other popular languages, including Java, Python, and C.

We will annotate our dataset with qualitative metrics to ensure we train with high-quality code. We will rely on various techniques to accurately label the dataset, including static code analysis tools, Stack Overflow discussions, and GitHub issue trackers. While mining GitHub repositories, strategies must be adopted to circumvent known pitfalls~\cite{mininggithub}. When extracting labels from Stack Overflow, we will address challenges posed by incomplete and unstructured code suggestions in the discussions by exploring automated program repair~\cite{programrepair}, including pattern-based~\cite{patternbasedrepair} and neural~\cite{neuralcoderepair} approaches.

\paragraph{Evaluation}

To evaluate the impact of the representative dataset, we will train code models like CodeBert~\cite{codebert}, AlphaCode~\cite{li2022competition}, and TravTrans~\cite{travtrans} on vanilla versions of CodeSearchNet. Subsequently, we will curate a variant of CodeSearchNet that represents real-world coding patterns and retrain the models on it. We then compare the differently trained model variants with each other. Standard metrics such as exact match score, statement-level accuracy, or BLEU~\cite{papineni_bleu_2002} are of limited value here, because programming tasks can be solved in multiple ways and different looking programs can be semantically equivalent~\cite{ren_codebleu_2020}. Therefore, we will evaluate the model performance using CodeBLEU~\cite{ren_codebleu_2020}, which also considers abstract syntax trees and data flow. Additionally, we will explore the use of techniques like semantic parsing and program analysis to estimate the semantic similarity of code segments.

\subsection{Capturing Code Structure and Semantics}

\paragraph{Challenge}

In Section~\ref{sec:intro}, we gave some examples of issues with state-of-the-art AI SD assistants, particularly their shortcomings in generating correct and secure code. Note that these are errors at the semantic level. How and to what extent code models can capture and respect program semantics remains an open research question\footnote{Earlier works have attempted to investigate this question for code transformers trained on abstract syntax tree (AST) representations~\cite{wan2022they}. However, such studies are limited, as ASTs may not capture program semantics very well, and the approaches cannot easily be adapted to new models.
}. One conceivable factor is that current code models treat code as formatted text and discard crucial semantic information such as control or data flow~\cite{ChakrabortyEtAl}, implicit usage constraints of third-party libraries, and domain-specific knowledge. It remains unclear what representations of code and SD knowledge should be used by these models.

\paragraph{Envisioned solution}

To quantify the problems with current textual code representations, we are developing an analysis technique to assess how much of a program's semantics transformer models learn to capture. We do this by calculating their \emph{semantic precision and recall}, i.e., how well the models' attention maps match the codes' control and data flow graphs. We will also consider the \emph{graph edit distance}~\cite{Gao2010} between them.

Further, since most semantic-rich representations of code are graph-based -- control flow graphs, data flow graphs, call graphs -- code models may benefit from native support of graph-based representations. To this end, we will explore different approaches and investigate their impact on the quality of suggested code: Firstly, we will work on flattening graph representations for code and using them as training data for sequence-based transformers. Positional encodings specialized for graph structures~\cite{shiv_novel_2019} may be instrumental in teaching the model how to interpret its input sequences correctly. Secondly, approaches related to graph attention networks~\cite{velickovic_graph_2018} and attention masking~\cite{mcdonald_syntax-based_2021} will be worth exploring, as they enable constraints on graph relations to be injected directly into the transformer's attention heads. Thirdly, we will explore the potential of graph neural networks~\cite{zhou_graph_2020} and graph transformers~\cite{graphtransformers} to operate directly on graph-based code representations. We will investigate the trade-offs between all three approaches. Potentially, this effort will lead to a foundational code model with increased semantic awareness.

\paragraph{Evaluation}

To evaluate the semantic awareness of our graph-based code models, we will use the novel analysis technique described above. Additionally, to evaluate their robustness, we randomly select code artifacts and make semantic-preserving changes to them -- renaming variables, translating for-statements into while-statements, etc. In a robust semantic-aware model, such semantics-preserving changes should typically not be reflected in large-scale changes to the attention maps. We will also compare the performance of the semantic-aware foundational code models we build with standard models like CodeBert, TravTrans, and GraphCodeBert~\cite{graphcodebert} on downstream tasks using the metrics mentioned in Subsection~\ref{subsec:dataset}.

\subsection{Code Quality}

\paragraph{Challenge}

Human software engineers receive feedback on their code as they write it through automated tools like compiler checks and linters, or from colleagues during pair programming and code reviews. Such feedback loops can have a very positive impact on code quality. For example, studies on pair programming report fewer bugs, better readability, higher test coverage and passing rates, and other benefits~\cite{pairprogramming1, pairprogramming2, pairprogramming3}. Code models, on the other hand, lack similar mechanisms. They are typically optimized for accuracy, not for qualitative criteria. We believe that a reliable AI SD assistant should provide code that not only accurately reflects existing code patterns, but is also of high quality according to well-established quality criteria and best practices.

\paragraph{Envisioned solution}

Inspired by the feedback loop described above, we propose an approach based on reinforcement learning (RL)~\cite{sutton2018reinforcement} to fine-tune code models for multiple code quality criteria. The approach shares ideas with actor--critic RL~\cite{konda_actor-critic_1999, sutton_policy_1999}, where, in our case, the code model takes the actor's role, and one or more code analysis tools serve as critics. Each critic evaluates the code generated by the actor and returns a token-wise reward value. The rewards are then aggregated (and traded off against each other) using \emph{utility functions} known from multi-objective RL~\cite{hayes_practical_2022}. Finally, the policy gradient is used to update the actor model, e.g., through proximal policy optimization (PPO)~\cite{schulman_proximal_2017}.

While ideas with actor--critic RL have recently been explored for program correctness~\cite{coderl}, our setting has a number of special characteristics that are not systematically explored: Since there is no universal metric for code quality, we resort to multiple critics, each specializing in one or more qualitative aspects. Possible critics for general best practices are linters and similar static analysis tools. To optimize security aspects, specialized security checkers like CogniCrypt~\cite{kruger_cognicrypt_2017} or CryptoGuard~\cite{rahaman_cryptoguard_2019} can be used. One could also consider more subjective criteria such as readability metrics~\cite{mainclassifier}. In our setting, critics must provide individual rewards for each token. This enables reward shaping and, thus, more efficient learning, as it conveys to the actor which actions (i.e., generated tokens) were particularly good or bad. Despite this, our setting has the benefit that we can generate a complete program before applying the critics to it. This allows us to use a wide variety of critics, even if they cannot handle partial (incomplete) programs. Lastly, we must point out that multi-objective RL with many (i.e., four or more) objectives is considered a hard problem~\cite{hayes_practical_2022}. In our case, however, the different code quality objectives can be assumed to be mostly non-conflicting. Therefore, there is no need to find all Pareto-optimal policies -- one is sufficient -- which allows us to use much more efficient algorithms.

\paragraph{Evaluation}

The effects of our code quality fine-tuning approach can be measured by a direct comparison of variants of a code model before and after fine-tuning. The same tools used as critics can be used to evaluate the code generated by the two variants. In addition, the code should be audited and rated by humans experts. For most code quality criteria, any coding benchmark is suitable, while specialized benchmarks like LLMSecEval~\cite{tony_llmseceval_2023} can be used to evaluate safety and security criteria.

\subsection{Explainability}

\paragraph{Challenge}

Current code models are not designed to base their answers on concrete background knowledge and use that knowledge to account for their answers to the user. While instruction fine-tuned (code) models~\cite{zhang_instruction_2023} often add explanations to their answers, these answers are the result of purely statistical computations and may be hallucinated. For instance, a recent study by Kabir et~al.~\cite{kabir_who_2023} found that ChatGPT answered Stack Overflow questions incorrectly half of the time. In our view, the issue of how to enable LLMs to provide correct and appropriate explanations for code remains unresolved\footnote{A closely related unresolved problem is how to keep a code model in sync with the ever-evolving correct coding patterns and best practices. This is particularly important when writing safety or security-critical software, where deprecated features should never be used. Given that expensive regular retraining is not desirable, explicit access to (up-to-date) background knowledge is required here as well. Hence, one approach may address both of these problems at once.}.

\paragraph{Envisioned solution}

For this challenge, we plan to investigate how to integrate \emph{code knowledge graphs} (CKGs) into our AI SD assistant. This way, we want to give the assistant access to a rich, easily updatable source of background knowledge that it can ground its generated code and explanations in. A potential subject for our investigation is IBM's \GraphGenForCode{}\footnote{\url{https://wala.github.io/graph4code/}}~\cite{abdelaziz2021graph4code}. This CKG is a set of triplets describing relationships (edges) between entities (nodes). For a given piece of code, the corresponding CKG contains nodes for each statement, with edges signifying control and data flow. Additionally, \GraphGenForCode{} has established initial semantic links between code fragments and corresponding forum discussions on Stack Overflow. However, the expressiveness of these name-based links is rather superficial.

To improve the expressiveness of the relationships between code and knowledge sources, we propose to enrich \GraphGenForCode{} with information about the task to which a particular code description or forum post relates. This information can be obtained using intent classification. Moreover, while Stack Overflow is a valuable source of discussions about and fixes to programming issues, the GitHub issue tracker can be another equally important source of such information. We thus propose to add links between code segments and GitHub issues to the CKG, along with additional information, such as whether a pull request successfully fixed the corresponding issue.

How to integrate the CKG into the assistant remains to be determined. One option is to augment the code model to query the CKG during the generation of its answer -- several papers have already demonstrated how to teach LLMs to use external tools or APIs~\cite{mialon2023augmented}. Alternatively, in a post-processing step, the CKG could be searched for code segments similar to the generated one. Then, the model could do a second pass over the generated code, using the retrieved information to ``decorate'' the generated code with explanations. In order to find relevant code samples in the CKG, a separate retriever model could be used~\cite{lu2022reacc}.

\paragraph{Evaluation}

To evaluate the extended CKG's ability to be integrated into the assistant and provide accurate and appropriate explanations to the software engineer, we will utilize evaluation protocols similar to those used to evaluate the original \GraphGenForCode{}.
We will involve human annotators and design an agreement metric to rank the quality of the assistant's suggestions and explanations. We will also evaluate the computing cost of enhancing large CKG's with complex semantic links and new sources of information such as GitHub issues.

\subsection{Controlled Code Generation}

\paragraph{Challenge}

All measures presented so far serve to enhance the capabilities of the AI SD assistant. However, due to the statistical nature of LLMs, even the best model may occasionally produce bad (e.g., incorrect, insecure, or low-quality) results. This lack of guarantees is a major hindrance to the trustworthiness of AI SD assistants.

\paragraph{Envisioned solution}

To provide the desired guarantees, we propose to equip the assistant with \emph{constrained (a.k.a. guided) decoding} (e.g.,~\cite{poesia_synchromesh_2022}) -- a recently popular extension of the decoding algorithms of causal LLMs that enforces compliance with certain rules for the generated text. At each generation step, all tokens that would lead to a violation of the rules are prevented from being generated by setting their probability to zero before applying the regular decoding strategy. The underlying rules can, e.g., be defined through regular expressions or grammars and the implementations can be very efficient~\cite{willard_efficient_2023}. This approach has already been applied to coding tasks, where it could, among other things, ensure syntactic correctness~\cite{poesia_synchromesh_2022} and prevent hallucinations~\cite{agrawal_guiding_2023}. Models equipped with constrained decoding often outperform larger models~\cite{agrawal_guiding_2023, poesia_synchromesh_2022, willard_efficient_2023}.

Since constrained decoding rules can be implemented and applied independently of each other, we envision a modular constrained decoding framework on top of the central code model. The modularity allows the user to select a ruleset appropriate for the current programming language and domain. There is a wide variety of rules that could be implemented. Constrained decoding is, however, no replacement for a capable code model for three reasons: rules must be checkable given only a program \emph{prefix}; for efficiency reasons, they should be \emph{incremental}, i.e., not require reanalysis of the program at each generation step; and, depending on the complexity of the rules, the constraints may be \emph{unsound} and/or \emph{incomplete}. Still, the lower bound on code quality, and thus the trustworthiness of the assistant, that constrained decoding provides is very valuable.

\paragraph{Evaluation}

Measuring how effective constrained decoding is in preventing undesired outputs is easy: First, we perform \emph{unconstrained} decoding on a series of prompts. Then we iterate over the output sequences using the constrained decoding framework and count in how many cases at least one of the rules was violated (and the constrained decoding would have intervened). This could also be insightful as it reveals typical errors made by unconstrained models. On the other hand, rules can be overly strict and guide the decoding into ``dead ends'' where the model can only terminate the generation or produce degenerate (i.e., valid but useless) results. How common this is compared to unconstrained decoding can be determined via standard benchmarks. The outcome of the described evaluations will of course highly depend on the used foundation model and ruleset.

\section{Future Plans}

The vision presented in this paper encompasses multiple ideas, all working towards creating a trustworthy AI software development assistant. Despite this, each idea is independent enough to form its own line of research. Therefore, they are worked on in parallel by different subgroups of researchers and lead to several full papers. At the current state, all components are conceptualized but still need to be implemented and evaluated as described above. Combining all outcomes to build the described programming assistant is a highly ambitious goal and will take several years to realize.

\section{Conclusion}

In this paper, we shared our understanding of the shortcomings of current code models and presented our vision of how they can be overcome. We considered aspects such as representative datasets, semantic code representations, code quality, explainability, and guarantees.

\begin{acks}
    Funded by the Hessian Ministry of Higher Education, Research, Science and the Arts within the cluster project \textit{The Third Wave of Artificial Intelligence} (3AI).
\end{acks}

\bibliographystyle{ACM-Reference-Format}
\bibliography{main}


\begin{thebibliography}{45}


\ifx \showCODEN    \undefined \def \showCODEN     #1{\unskip}     \fi
\ifx \showDOI      \undefined \def \showDOI       #1{#1}\fi
\ifx \showISBNx    \undefined \def \showISBNx     #1{\unskip}     \fi
\ifx \showISBNxiii \undefined \def \showISBNxiii  #1{\unskip}     \fi
\ifx \showISSN     \undefined \def \showISSN      #1{\unskip}     \fi
\ifx \showLCCN     \undefined \def \showLCCN      #1{\unskip}     \fi
\ifx \shownote     \undefined \def \shownote      #1{#1}          \fi
\ifx \showarticletitle \undefined \def \showarticletitle #1{#1}   \fi
\ifx \showURL      \undefined \def \showURL       {\relax}        \fi
\providecommand\bibfield[2]{#2}
\providecommand\bibinfo[2]{#2}
\providecommand\natexlab[1]{#1}
\providecommand\showeprint[2][]{arXiv:#2}

\bibitem[Abdelaziz et~al\mbox{.}(2021)]%
        {abdelaziz2021graph4code}
\bibfield{author}{\bibinfo{person}{Ibrahim Abdelaziz}, \bibinfo{person}{Julian
  Dolby}, \bibinfo{person}{Jamie~P. McCusker}, {and} \bibinfo{person}{Kavitha
  Srinivas}.} \bibinfo{year}{2021}\natexlab{}.
\newblock \showarticletitle{A Toolkit for Generating Code Knowledge Graphs}. In
  \bibinfo{booktitle}{\emph{{K-CAP} '21: Knowledge Capture Conference, Virtual
  Event, USA, December 2-3, 2021}},
  \bibfield{editor}{\bibinfo{person}{Anna~Lisa Gentile} {and}
  \bibinfo{person}{Rafael Gon{\c{c}}alves}} (Eds.). \bibinfo{publisher}{{ACM}},
  \bibinfo{pages}{137--144}.
\newblock


\bibitem[Agrawal et~al\mbox{.}(2023)]%
        {agrawal_guiding_2023}
\bibfield{author}{\bibinfo{person}{Lakshya~A Agrawal}, \bibinfo{person}{Aditya
  Kanade}, \bibinfo{person}{Navin Goyal}, \bibinfo{person}{Shuvendu~K. Lahiri},
  {and} \bibinfo{person}{Sriram~K. Rajamani}.} \bibinfo{year}{2023}\natexlab{}.
\newblock \showarticletitle{Guiding Language Models of Code with Global Context
  using Monitors}.
\newblock \bibinfo{journal}{\emph{CoRR}}  \bibinfo{volume}{abs/2306.10763}
  (\bibinfo{year}{2023}).
\newblock


\bibitem[Boehm et~al\mbox{.}(2000)]%
        {boehm2000software}
\bibfield{author}{\bibinfo{person}{Barry~W. Boehm}, \bibinfo{person}{Chris
  Abts}, {and} \bibinfo{person}{Sunita Chulani}.}
  \bibinfo{year}{2000}\natexlab{}.
\newblock \showarticletitle{Software development cost estimation approaches -
  {A} survey}.
\newblock \bibinfo{journal}{\emph{Ann. Softw. Eng.}}  \bibinfo{volume}{10}
  (\bibinfo{year}{2000}), \bibinfo{pages}{177--205}.
\newblock


\bibitem[Chakraborty et~al\mbox{.}(2022)]%
        {ChakrabortyEtAl}
\bibfield{author}{\bibinfo{person}{Saikat Chakraborty}, \bibinfo{person}{Rahul
  Krishna}, \bibinfo{person}{Yangruibo Ding}, {and} \bibinfo{person}{Baishakhi
  Ray}.} \bibinfo{year}{2022}\natexlab{}.
\newblock \showarticletitle{Deep Learning Based Vulnerability Detection: Are We
  There Yet?}
\newblock \bibinfo{journal}{\emph{{IEEE} Trans. Software Eng.}}
  \bibinfo{volume}{48}, \bibinfo{number}{9} (\bibinfo{year}{2022}),
  \bibinfo{pages}{3280--3296}.
\newblock


\bibitem[Feng et~al\mbox{.}(2020)]%
        {codebert}
\bibfield{author}{\bibinfo{person}{Zhangyin Feng}, \bibinfo{person}{Daya Guo},
  \bibinfo{person}{Duyu Tang}, \bibinfo{person}{Nan Duan},
  \bibinfo{person}{Xiaocheng Feng}, \bibinfo{person}{Ming Gong},
  \bibinfo{person}{Linjun Shou}, \bibinfo{person}{Bing Qin},
  \bibinfo{person}{Ting Liu}, \bibinfo{person}{Daxin Jiang}, {and}
  \bibinfo{person}{Ming Zhou}.} \bibinfo{year}{2020}\natexlab{}.
\newblock \showarticletitle{CodeBERT: {A} Pre-Trained Model for Programming and
  Natural Languages}. In \bibinfo{booktitle}{\emph{Findings of the Association
  for Computational Linguistics: {EMNLP} 2020, Online Event, 16-20 November
  2020}} \emph{(\bibinfo{series}{Findings of {ACL}},
  Vol.~\bibinfo{volume}{{EMNLP} 2020})},
  \bibfield{editor}{\bibinfo{person}{Trevor Cohn}, \bibinfo{person}{Yulan He},
  {and} \bibinfo{person}{Yang Liu}} (Eds.). \bibinfo{publisher}{Association for
  Computational Linguistics}, \bibinfo{pages}{1536--1547}.
\newblock


\bibitem[Gao et~al\mbox{.}(2010)]%
        {Gao2010}
\bibfield{author}{\bibinfo{person}{Xinbo Gao}, \bibinfo{person}{Bing Xiao},
  \bibinfo{person}{Dacheng Tao}, {and} \bibinfo{person}{Xuelong Li}.}
  \bibinfo{year}{2010}\natexlab{}.
\newblock \showarticletitle{A survey of graph edit distance}.
\newblock \bibinfo{journal}{\emph{Pattern Anal. Appl.}} \bibinfo{volume}{13},
  \bibinfo{number}{1} (\bibinfo{year}{2010}), \bibinfo{pages}{113--129}.
\newblock


\bibitem[Goues et~al\mbox{.}(2019)]%
        {programrepair}
\bibfield{author}{\bibinfo{person}{Claire~Le Goues}, \bibinfo{person}{Michael
  Pradel}, {and} \bibinfo{person}{Abhik Roychoudhury}.}
  \bibinfo{year}{2019}\natexlab{}.
\newblock \showarticletitle{Automated program repair}.
\newblock \bibinfo{journal}{\emph{Commun. {ACM}}} \bibinfo{volume}{62},
  \bibinfo{number}{12} (\bibinfo{year}{2019}), \bibinfo{pages}{56--65}.
\newblock


\bibitem[Guo et~al\mbox{.}(2021)]%
        {graphcodebert}
\bibfield{author}{\bibinfo{person}{Daya Guo}, \bibinfo{person}{Shuo Ren},
  \bibinfo{person}{Shuai Lu}, \bibinfo{person}{Zhangyin Feng},
  \bibinfo{person}{Duyu Tang}, \bibinfo{person}{Shujie Liu},
  \bibinfo{person}{Long Zhou}, \bibinfo{person}{Nan Duan},
  \bibinfo{person}{Alexey Svyatkovskiy}, \bibinfo{person}{Shengyu Fu},
  \bibinfo{person}{Michele Tufano}, \bibinfo{person}{Shao~Kun Deng},
  \bibinfo{person}{Colin~B. Clement}, \bibinfo{person}{Dawn Drain},
  \bibinfo{person}{Neel Sundaresan}, \bibinfo{person}{Jian Yin},
  \bibinfo{person}{Daxin Jiang}, {and} \bibinfo{person}{Ming Zhou}.}
  \bibinfo{year}{2021}\natexlab{}.
\newblock \showarticletitle{GraphCodeBERT: Pre-training Code Representations
  with Data Flow}. In \bibinfo{booktitle}{\emph{9th International Conference on
  Learning Representations, {ICLR} 2021, Virtual Event, Austria, May 3-7,
  2021}}. \bibinfo{publisher}{OpenReview.net}.
\newblock


\bibitem[Gupta et~al\mbox{.}(2017)]%
        {neuralcoderepair}
\bibfield{author}{\bibinfo{person}{Rahul Gupta}, \bibinfo{person}{Soham Pal},
  \bibinfo{person}{Aditya Kanade}, {and} \bibinfo{person}{Shirish~K. Shevade}.}
  \bibinfo{year}{2017}\natexlab{}.
\newblock \showarticletitle{DeepFix: Fixing Common {C} Language Errors by Deep
  Learning}. In \bibinfo{booktitle}{\emph{Proceedings of the Thirty-First
  {AAAI} Conference on Artificial Intelligence, February 4-9, 2017, San
  Francisco, California, {USA}}}, \bibfield{editor}{\bibinfo{person}{Satinder
  Singh} {and} \bibinfo{person}{Shaul Markovitch}} (Eds.).
  \bibinfo{publisher}{{AAAI} Press}, \bibinfo{pages}{1345--1351}.
\newblock


\bibitem[Hayes et~al\mbox{.}(2022)]%
        {hayes_practical_2022}
\bibfield{author}{\bibinfo{person}{Conor~F. Hayes}, \bibinfo{person}{Roxana
  Radulescu}, \bibinfo{person}{Eugenio Bargiacchi}, \bibinfo{person}{Johan
  K{\"{a}}llstr{\"{o}}m}, \bibinfo{person}{Matthew Macfarlane},
  \bibinfo{person}{Mathieu Reymond}, \bibinfo{person}{Timothy Verstraeten},
  \bibinfo{person}{Luisa~M. Zintgraf}, \bibinfo{person}{Richard Dazeley},
  \bibinfo{person}{Fredrik Heintz}, \bibinfo{person}{Enda Howley},
  \bibinfo{person}{Athirai~A. Irissappane}, \bibinfo{person}{Patrick Mannion},
  \bibinfo{person}{Ann Now{\'{e}}}, \bibinfo{person}{Gabriel de
  Oliveira~Ramos}, \bibinfo{person}{Marcello Restelli}, \bibinfo{person}{Peter
  Vamplew}, {and} \bibinfo{person}{Diederik~M. Roijers}.}
  \bibinfo{year}{2022}\natexlab{}.
\newblock \showarticletitle{A practical guide to multi-objective reinforcement
  learning and planning}.
\newblock \bibinfo{journal}{\emph{Auton. Agents Multi Agent Syst.}}
  \bibinfo{volume}{36}, \bibinfo{number}{1} (\bibinfo{year}{2022}),
  \bibinfo{pages}{26}.
\newblock


\bibitem[Hellendoorn et~al\mbox{.}(2019)]%
        {whenfail}
\bibfield{author}{\bibinfo{person}{Vincent~J. Hellendoorn},
  \bibinfo{person}{Sebastian Proksch}, \bibinfo{person}{Harald~C. Gall}, {and}
  \bibinfo{person}{Alberto Bacchelli}.} \bibinfo{year}{2019}\natexlab{}.
\newblock \showarticletitle{When code completion fails: a case study on
  real-world completions}. In \bibinfo{booktitle}{\emph{Proceedings of the 41st
  International Conference on Software Engineering, {ICSE} 2019, Montreal, QC,
  Canada, May 25-31, 2019}}, \bibfield{editor}{\bibinfo{person}{Joanne~M.
  Atlee}, \bibinfo{person}{Tevfik Bultan}, {and} \bibinfo{person}{Jon Whittle}}
  (Eds.). \bibinfo{publisher}{{IEEE} / {ACM}}, \bibinfo{pages}{960--970}.
\newblock


\bibitem[Husain et~al\mbox{.}(2019)]%
        {husain2019codesearchnet}
\bibfield{author}{\bibinfo{person}{Hamel Husain}, \bibinfo{person}{Ho{-}Hsiang
  Wu}, \bibinfo{person}{Tiferet Gazit}, \bibinfo{person}{Miltiadis Allamanis},
  {and} \bibinfo{person}{Marc Brockschmidt}.} \bibinfo{year}{2019}\natexlab{}.
\newblock \showarticletitle{CodeSearchNet Challenge: Evaluating the State of
  Semantic Code Search}.
\newblock \bibinfo{journal}{\emph{CoRR}}  \bibinfo{volume}{abs/1909.09436}
  (\bibinfo{year}{2019}).
\newblock


\bibitem[Kabir et~al\mbox{.}(2023)]%
        {kabir_who_2023}
\bibfield{author}{\bibinfo{person}{Samia Kabir}, \bibinfo{person}{David~N.
  Udo{-}Imeh}, \bibinfo{person}{Bonan Kou}, {and} \bibinfo{person}{Tianyi
  Zhang}.} \bibinfo{year}{2023}\natexlab{}.
\newblock \showarticletitle{Who Answers It Better? An In-Depth Analysis of
  ChatGPT and Stack Overflow Answers to Software Engineering Questions}.
\newblock \bibinfo{journal}{\emph{CoRR}}  \bibinfo{volume}{abs/2308.02312}
  (\bibinfo{year}{2023}).
\newblock


\bibitem[Kalliamvakou et~al\mbox{.}(2014)]%
        {mininggithub}
\bibfield{author}{\bibinfo{person}{Eirini Kalliamvakou},
  \bibinfo{person}{Georgios Gousios}, \bibinfo{person}{Kelly Blincoe},
  \bibinfo{person}{Leif Singer}, \bibinfo{person}{Daniel~M. Germ{\'{a}}n},
  {and} \bibinfo{person}{Daniela~E. Damian}.} \bibinfo{year}{2014}\natexlab{}.
\newblock \showarticletitle{The promises and perils of mining GitHub}. In
  \bibinfo{booktitle}{\emph{11th Working Conference on Mining Software
  Repositories, {MSR} 2014, Proceedings, May 31 - June 1, 2014, Hyderabad,
  India}}, \bibfield{editor}{\bibinfo{person}{Premkumar~T. Devanbu},
  \bibinfo{person}{Sung Kim}, {and} \bibinfo{person}{Martin Pinzger}} (Eds.).
  \bibinfo{publisher}{{ACM}}, \bibinfo{pages}{92--101}.
\newblock


\bibitem[Kim et~al\mbox{.}(2013)]%
        {patternbasedrepair}
\bibfield{author}{\bibinfo{person}{Dongsun Kim}, \bibinfo{person}{Jaechang
  Nam}, \bibinfo{person}{Jaewoo Song}, {and} \bibinfo{person}{Sunghun Kim}.}
  \bibinfo{year}{2013}\natexlab{}.
\newblock \showarticletitle{Automatic patch generation learned from
  human-written patches}. In \bibinfo{booktitle}{\emph{35th International
  Conference on Software Engineering, {ICSE} '13, San Francisco, CA, USA, May
  18-26, 2013}}, \bibfield{editor}{\bibinfo{person}{David Notkin},
  \bibinfo{person}{Betty H.~C. Cheng}, {and} \bibinfo{person}{Klaus Pohl}}
  (Eds.). \bibinfo{publisher}{{IEEE} Computer Society},
  \bibinfo{pages}{802--811}.
\newblock


\bibitem[Kim et~al\mbox{.}(2021)]%
        {travtrans}
\bibfield{author}{\bibinfo{person}{Seohyun Kim}, \bibinfo{person}{Jinman Zhao},
  \bibinfo{person}{Yuchi Tian}, {and} \bibinfo{person}{Satish Chandra}.}
  \bibinfo{year}{2021}\natexlab{}.
\newblock \showarticletitle{Code Prediction by Feeding Trees to Transformers}.
  In \bibinfo{booktitle}{\emph{43rd {IEEE/ACM} International Conference on
  Software Engineering, {ICSE} 2021, Madrid, Spain, 22-30 May 2021}}.
  \bibinfo{publisher}{{IEEE}}, \bibinfo{pages}{150--162}.
\newblock


\bibitem[Konda and Tsitsiklis(1999)]%
        {konda_actor-critic_1999}
\bibfield{author}{\bibinfo{person}{Vijay~R. Konda} {and}
  \bibinfo{person}{John~N. Tsitsiklis}.} \bibinfo{year}{1999}\natexlab{}.
\newblock \showarticletitle{Actor-Critic Algorithms}. In
  \bibinfo{booktitle}{\emph{Advances in Neural Information Processing Systems
  12, {[NIPS} Conference, Denver, Colorado, USA, November 29 - December 4,
  1999]}}, \bibfield{editor}{\bibinfo{person}{Sara~A. Solla},
  \bibinfo{person}{Todd~K. Leen}, {and} \bibinfo{person}{Klaus{-}Robert
  M{\"{u}}ller}} (Eds.). \bibinfo{publisher}{The {MIT} Press},
  \bibinfo{pages}{1008--1014}.
\newblock


\bibitem[Kr{\"{u}}ger et~al\mbox{.}(2017)]%
        {kruger_cognicrypt_2017}
\bibfield{author}{\bibinfo{person}{Stefan Kr{\"{u}}ger}, \bibinfo{person}{Sarah
  Nadi}, \bibinfo{person}{Michael Reif}, \bibinfo{person}{Karim Ali},
  \bibinfo{person}{Mira Mezini}, \bibinfo{person}{Eric Bodden},
  \bibinfo{person}{Florian G{\"{o}}pfert}, \bibinfo{person}{Felix
  G{\"{u}}nther}, \bibinfo{person}{Christian Weinert}, \bibinfo{person}{Daniel
  Demmler}, {and} \bibinfo{person}{Ram Kamath}.}
  \bibinfo{year}{2017}\natexlab{}.
\newblock \showarticletitle{CogniCrypt: supporting developers in using
  cryptography}. In \bibinfo{booktitle}{\emph{Proceedings of the 32nd
  {IEEE/ACM} International Conference on Automated Software Engineering, {ASE}
  2017, Urbana, IL, USA, October 30 - November 03, 2017}},
  \bibfield{editor}{\bibinfo{person}{Grigore Rosu},
  \bibinfo{person}{Massimiliano~Di Penta}, {and} \bibinfo{person}{Tien~N.
  Nguyen}} (Eds.). \bibinfo{publisher}{{IEEE} Computer Society},
  \bibinfo{pages}{931--936}.
\newblock


\bibitem[Le et~al\mbox{.}(2022)]%
        {coderl}
\bibfield{author}{\bibinfo{person}{Hung Le}, \bibinfo{person}{Yue Wang},
  \bibinfo{person}{Akhilesh~Deepak Gotmare}, \bibinfo{person}{Silvio Savarese},
  {and} \bibinfo{person}{Steven~Chu{-}Hong Hoi}.}
  \bibinfo{year}{2022}\natexlab{}.
\newblock \showarticletitle{CodeRL: Mastering Code Generation through
  Pretrained Models and Deep Reinforcement Learning}. In
  \bibinfo{booktitle}{\emph{NeurIPS}}.
\newblock


\bibitem[Li et~al\mbox{.}(2022)]%
        {li2022competition}
\bibfield{author}{\bibinfo{person}{Yujia Li}, \bibinfo{person}{David~H. Choi},
  \bibinfo{person}{Junyoung Chung}, \bibinfo{person}{Nate Kushman},
  \bibinfo{person}{Julian Schrittwieser}, \bibinfo{person}{R{\'{e}}mi Leblond},
  \bibinfo{person}{Tom Eccles}, \bibinfo{person}{James Keeling},
  \bibinfo{person}{Felix Gimeno}, \bibinfo{person}{Agustin~Dal Lago},
  \bibinfo{person}{Thomas Hubert}, \bibinfo{person}{Peter Choy},
  \bibinfo{person}{Cyprien de Masson~d'Autume}, \bibinfo{person}{Igor
  Babuschkin}, \bibinfo{person}{Xinyun Chen}, \bibinfo{person}{Po{-}Sen Huang},
  \bibinfo{person}{Johannes Welbl}, \bibinfo{person}{Sven Gowal},
  \bibinfo{person}{Alexey Cherepanov}, \bibinfo{person}{James Molloy},
  \bibinfo{person}{Daniel~J. Mankowitz}, \bibinfo{person}{Esme~Sutherland
  Robson}, \bibinfo{person}{Pushmeet Kohli}, \bibinfo{person}{Nando de
  Freitas}, \bibinfo{person}{Koray Kavukcuoglu}, {and} \bibinfo{person}{Oriol
  Vinyals}.} \bibinfo{year}{2022}\natexlab{}.
\newblock \showarticletitle{Competition-Level Code Generation with AlphaCode}.
\newblock \bibinfo{journal}{\emph{CoRR}}  \bibinfo{volume}{abs/2203.07814}
  (\bibinfo{year}{2022}).
\newblock


\bibitem[Lu et~al\mbox{.}(2022)]%
        {lu2022reacc}
\bibfield{author}{\bibinfo{person}{Shuai Lu}, \bibinfo{person}{Nan Duan},
  \bibinfo{person}{Hojae Han}, \bibinfo{person}{Daya Guo},
  \bibinfo{person}{Seung{-}won Hwang}, {and} \bibinfo{person}{Alexey
  Svyatkovskiy}.} \bibinfo{year}{2022}\natexlab{}.
\newblock \showarticletitle{ReACC: {A} Retrieval-Augmented Code Completion
  Framework}. In \bibinfo{booktitle}{\emph{Proceedings of the 60th Annual
  Meeting of the Association for Computational Linguistics (Volume 1: Long
  Papers), {ACL} 2022, Dublin, Ireland, May 22-27, 2022}},
  \bibfield{editor}{\bibinfo{person}{Smaranda Muresan},
  \bibinfo{person}{Preslav Nakov}, {and} \bibinfo{person}{Aline Villavicencio}}
  (Eds.). \bibinfo{publisher}{Association for Computational Linguistics},
  \bibinfo{pages}{6227--6240}.
\newblock


\bibitem[McDonald and Chiang(2021)]%
        {mcdonald_syntax-based_2021}
\bibfield{author}{\bibinfo{person}{Colin McDonald} {and} \bibinfo{person}{David
  Chiang}.} \bibinfo{year}{2021}\natexlab{}.
\newblock \showarticletitle{Syntax-Based Attention Masking for Neural Machine
  Translation}. In \bibinfo{booktitle}{\emph{Proceedings of the 2021 Conference
  of the North American Chapter of the Association for Computational
  Linguistics: Student Research Workshop, {NAACL-HLT} 2021, Online, June 6-11,
  2021}}, \bibfield{editor}{\bibinfo{person}{Esin Durmus},
  \bibinfo{person}{Vivek Gupta}, \bibinfo{person}{Nelson Liu},
  \bibinfo{person}{Nanyun Peng}, {and} \bibinfo{person}{Yu~Su}} (Eds.).
  \bibinfo{publisher}{Association for Computational Linguistics},
  \bibinfo{pages}{47--52}.
\newblock


\bibitem[Mialon et~al\mbox{.}(2023)]%
        {mialon2023augmented}
\bibfield{author}{\bibinfo{person}{Gr{\'{e}}goire Mialon},
  \bibinfo{person}{Roberto Dess{\`{\i}}}, \bibinfo{person}{Maria Lomeli},
  \bibinfo{person}{Christoforos Nalmpantis}, \bibinfo{person}{Ramakanth
  Pasunuru}, \bibinfo{person}{Roberta Raileanu}, \bibinfo{person}{Baptiste
  Rozi{\`{e}}re}, \bibinfo{person}{Timo Schick}, \bibinfo{person}{Jane
  Dwivedi{-}Yu}, \bibinfo{person}{Asli Celikyilmaz}, \bibinfo{person}{Edouard
  Grave}, \bibinfo{person}{Yann LeCun}, {and} \bibinfo{person}{Thomas
  Scialom}.} \bibinfo{year}{2023}\natexlab{}.
\newblock \showarticletitle{Augmented Language Models: a Survey}.
\newblock \bibinfo{journal}{\emph{CoRR}}  \bibinfo{volume}{abs/2302.07842}
  (\bibinfo{year}{2023}).
\newblock


\bibitem[Nadi et~al\mbox{.}(2016)]%
        {nadi2016jumping}
\bibfield{author}{\bibinfo{person}{Sarah Nadi}, \bibinfo{person}{Stefan
  Kr{\"{u}}ger}, \bibinfo{person}{Mira Mezini}, {and} \bibinfo{person}{Eric
  Bodden}.} \bibinfo{year}{2016}\natexlab{}.
\newblock \showarticletitle{Jumping through hoops: why do Java developers
  struggle with cryptography APIs?}. In \bibinfo{booktitle}{\emph{Proceedings
  of the 38th International Conference on Software Engineering, {ICSE} 2016,
  Austin, TX, USA, May 14-22, 2016}},
  \bibfield{editor}{\bibinfo{person}{Laura~K. Dillon}, \bibinfo{person}{Willem
  Visser}, {and} \bibinfo{person}{Laurie~A. Williams}} (Eds.).
  \bibinfo{publisher}{{ACM}}, \bibinfo{pages}{935--946}.
\newblock


\bibitem[Nawrocki and Wojciechowski(2001)]%
        {pairprogramming3}
\bibfield{author}{\bibinfo{person}{Jerzy Nawrocki} {and} \bibinfo{person}{Adam
  Wojciechowski}.} \bibinfo{year}{2001}\natexlab{}.
\newblock \showarticletitle{Experimental evaluation of pair programming}.
\newblock \bibinfo{journal}{\emph{European Software Control and Metrics
  (Escom)}} (\bibinfo{year}{2001}), \bibinfo{pages}{99--101}.
\newblock


\bibitem[Papineni et~al\mbox{.}(2002)]%
        {papineni_bleu_2002}
\bibfield{author}{\bibinfo{person}{Kishore Papineni}, \bibinfo{person}{Salim
  Roukos}, \bibinfo{person}{Todd Ward}, {and} \bibinfo{person}{Wei{-}Jing
  Zhu}.} \bibinfo{year}{2002}\natexlab{}.
\newblock \showarticletitle{Bleu: a Method for Automatic Evaluation of Machine
  Translation}. In \bibinfo{booktitle}{\emph{Proceedings of the 40th Annual
  Meeting of the Association for Computational Linguistics, July 6-12, 2002,
  Philadelphia, PA, {USA}}}. \bibinfo{publisher}{{ACL}},
  \bibinfo{pages}{311--318}.
\newblock


\bibitem[Pearce et~al\mbox{.}(2022)]%
        {copilotstudy40}
\bibfield{author}{\bibinfo{person}{Hammond Pearce}, \bibinfo{person}{Baleegh
  Ahmad}, \bibinfo{person}{Benjamin Tan}, \bibinfo{person}{Brendan
  Dolan{-}Gavitt}, {and} \bibinfo{person}{Ramesh Karri}.}
  \bibinfo{year}{2022}\natexlab{}.
\newblock \showarticletitle{Asleep at the Keyboard? Assessing the Security of
  GitHub Copilot's Code Contributions}. In \bibinfo{booktitle}{\emph{43rd
  {IEEE} Symposium on Security and Privacy, {SP} 2022, San Francisco, CA, USA,
  May 22-26, 2022}}. \bibinfo{publisher}{{IEEE}}, \bibinfo{pages}{754--768}.
\newblock


\bibitem[Perry et~al\mbox{.}(2023)]%
        {perry2022users}
\bibfield{author}{\bibinfo{person}{Neil Perry}, \bibinfo{person}{Megha
  Srivastava}, \bibinfo{person}{Deepak Kumar}, {and} \bibinfo{person}{Dan
  Boneh}.} \bibinfo{year}{2023}\natexlab{}.
\newblock \showarticletitle{Do Users Write More Insecure Code with {AI}
  Assistants?}. In \bibinfo{booktitle}{\emph{Proceedings of the 2023 {ACM}
  {SIGSAC} Conference on Computer and Communications Security, {CCS} 2023,
  Copenhagen, Denmark, November 26-30, 2023}},
  \bibfield{editor}{\bibinfo{person}{Weizhi Meng},
  \bibinfo{person}{Christian~Damsgaard Jensen}, \bibinfo{person}{Cas Cremers},
  {and} \bibinfo{person}{Engin Kirda}} (Eds.). \bibinfo{publisher}{{ACM}},
  \bibinfo{pages}{2785--2799}.
\newblock


\bibitem[Poesia et~al\mbox{.}(2022)]%
        {poesia_synchromesh_2022}
\bibfield{author}{\bibinfo{person}{Gabriel Poesia}, \bibinfo{person}{Alex
  Polozov}, \bibinfo{person}{Vu Le}, \bibinfo{person}{Ashish Tiwari},
  \bibinfo{person}{Gustavo Soares}, \bibinfo{person}{Christopher Meek}, {and}
  \bibinfo{person}{Sumit Gulwani}.} \bibinfo{year}{2022}\natexlab{}.
\newblock \showarticletitle{Synchromesh: Reliable Code Generation from
  Pre-trained Language Models}. In \bibinfo{booktitle}{\emph{The Tenth
  International Conference on Learning Representations, {ICLR} 2022, Virtual
  Event, April 25-29, 2022}}. \bibinfo{publisher}{OpenReview.net}.
\newblock


\bibitem[Rahaman et~al\mbox{.}(2019)]%
        {rahaman_cryptoguard_2019}
\bibfield{author}{\bibinfo{person}{Sazzadur Rahaman}, \bibinfo{person}{Ya
  Xiao}, \bibinfo{person}{Sharmin Afrose}, \bibinfo{person}{Fahad Shaon},
  \bibinfo{person}{Ke Tian}, \bibinfo{person}{Miles Frantz},
  \bibinfo{person}{Murat Kantarcioglu}, {and} \bibinfo{person}{Danfeng~(Daphne)
  Yao}.} \bibinfo{year}{2019}\natexlab{}.
\newblock \showarticletitle{CryptoGuard: High Precision Detection of
  Cryptographic Vulnerabilities in Massive-sized Java Projects}. In
  \bibinfo{booktitle}{\emph{Proceedings of the 2019 {ACM} {SIGSAC} Conference
  on Computer and Communications Security, {CCS} 2019, London, UK, November
  11-15, 2019}}, \bibfield{editor}{\bibinfo{person}{Lorenzo Cavallaro},
  \bibinfo{person}{Johannes Kinder}, \bibinfo{person}{XiaoFeng Wang}, {and}
  \bibinfo{person}{Jonathan Katz}} (Eds.). \bibinfo{publisher}{{ACM}},
  \bibinfo{pages}{2455--2472}.
\newblock


\bibitem[Ren et~al\mbox{.}(2020)]%
        {ren_codebleu_2020}
\bibfield{author}{\bibinfo{person}{Shuo Ren}, \bibinfo{person}{Daya Guo},
  \bibinfo{person}{Shuai Lu}, \bibinfo{person}{Long Zhou},
  \bibinfo{person}{Shujie Liu}, \bibinfo{person}{Duyu Tang},
  \bibinfo{person}{Neel Sundaresan}, \bibinfo{person}{Ming Zhou},
  \bibinfo{person}{Ambrosio Blanco}, {and} \bibinfo{person}{Shuai Ma}.}
  \bibinfo{year}{2020}\natexlab{}.
\newblock \showarticletitle{CodeBLEU: a Method for Automatic Evaluation of Code
  Synthesis}.
\newblock \bibinfo{journal}{\emph{CoRR}}  \bibinfo{volume}{abs/2009.10297}
  (\bibinfo{year}{2020}).
\newblock


\bibitem[Scalabrino et~al\mbox{.}(2018)]%
        {mainclassifier}
\bibfield{author}{\bibinfo{person}{Simone Scalabrino}, \bibinfo{person}{Mario
  Linares{-}V{\'{a}}squez}, \bibinfo{person}{Rocco Oliveto}, {and}
  \bibinfo{person}{Denys Poshyvanyk}.} \bibinfo{year}{2018}\natexlab{}.
\newblock \showarticletitle{A comprehensive model for code readability}.
\newblock \bibinfo{journal}{\emph{J. Softw. Evol. Process.}}
  \bibinfo{volume}{30}, \bibinfo{number}{6} (\bibinfo{year}{2018}).
\newblock


\bibitem[Schulman et~al\mbox{.}(2017)]%
        {schulman_proximal_2017}
\bibfield{author}{\bibinfo{person}{John Schulman}, \bibinfo{person}{Filip
  Wolski}, \bibinfo{person}{Prafulla Dhariwal}, \bibinfo{person}{Alec Radford},
  {and} \bibinfo{person}{Oleg Klimov}.} \bibinfo{year}{2017}\natexlab{}.
\newblock \showarticletitle{Proximal Policy Optimization Algorithms}.
\newblock \bibinfo{journal}{\emph{CoRR}}  \bibinfo{volume}{abs/1707.06347}
  (\bibinfo{year}{2017}).
\newblock


\bibitem[Shiv and Quirk(2019)]%
        {shiv_novel_2019}
\bibfield{author}{\bibinfo{person}{Vighnesh~Leonardo Shiv} {and}
  \bibinfo{person}{Chris Quirk}.} \bibinfo{year}{2019}\natexlab{}.
\newblock \showarticletitle{Novel positional encodings to enable tree-based
  transformers}. In \bibinfo{booktitle}{\emph{Advances in Neural Information
  Processing Systems 32: Annual Conference on Neural Information Processing
  Systems 2019, NeurIPS 2019, December 8-14, 2019, Vancouver, BC, Canada}},
  \bibfield{editor}{\bibinfo{person}{Hanna~M. Wallach}, \bibinfo{person}{Hugo
  Larochelle}, \bibinfo{person}{Alina Beygelzimer}, \bibinfo{person}{Florence
  d'Alch{\'{e}}{-}Buc}, \bibinfo{person}{Emily~B. Fox}, {and}
  \bibinfo{person}{Roman Garnett}} (Eds.). \bibinfo{pages}{12058--12068}.
\newblock


\bibitem[Sutton and Barto(1998)]%
        {sutton2018reinforcement}
\bibfield{author}{\bibinfo{person}{Richard~S. Sutton} {and}
  \bibinfo{person}{Andrew~G. Barto}.} \bibinfo{year}{1998}\natexlab{}.
\newblock \bibinfo{booktitle}{\emph{Reinforcement learning - an introduction}}.
\newblock \bibinfo{publisher}{{MIT} Press}.
\newblock
\showISBNx{978-0-262-19398-6}


\bibitem[Sutton et~al\mbox{.}(1999)]%
        {sutton_policy_1999}
\bibfield{author}{\bibinfo{person}{Richard~S. Sutton},
  \bibinfo{person}{David~A. McAllester}, \bibinfo{person}{Satinder Singh},
  {and} \bibinfo{person}{Yishay Mansour}.} \bibinfo{year}{1999}\natexlab{}.
\newblock \showarticletitle{Policy Gradient Methods for Reinforcement Learning
  with Function Approximation}. In \bibinfo{booktitle}{\emph{Advances in Neural
  Information Processing Systems 12, {[NIPS} Conference, Denver, Colorado, USA,
  November 29 - December 4, 1999]}}, \bibfield{editor}{\bibinfo{person}{Sara~A.
  Solla}, \bibinfo{person}{Todd~K. Leen}, {and} \bibinfo{person}{Klaus{-}Robert
  M{\"{u}}ller}} (Eds.). \bibinfo{publisher}{The {MIT} Press},
  \bibinfo{pages}{1057--1063}.
\newblock


\bibitem[Tony et~al\mbox{.}(2023)]%
        {tony_llmseceval_2023}
\bibfield{author}{\bibinfo{person}{Catherine Tony}, \bibinfo{person}{Markus
  Mutas}, \bibinfo{person}{Nicol{\'{a}}s E.~D{\'{\i}}az Ferreyra}, {and}
  \bibinfo{person}{Riccardo Scandariato}.} \bibinfo{year}{2023}\natexlab{}.
\newblock \showarticletitle{LLMSecEval: {A} Dataset of Natural Language Prompts
  for Security Evaluations}. In \bibinfo{booktitle}{\emph{20th {IEEE/ACM}
  International Conference on Mining Software Repositories, {MSR} 2023,
  Melbourne, Australia, May 15-16, 2023}}. \bibinfo{publisher}{{IEEE}},
  \bibinfo{pages}{588--592}.
\newblock


\bibitem[Velickovic et~al\mbox{.}(2018)]%
        {velickovic_graph_2018}
\bibfield{author}{\bibinfo{person}{Petar Velickovic}, \bibinfo{person}{Guillem
  Cucurull}, \bibinfo{person}{Arantxa Casanova}, \bibinfo{person}{Adriana
  Romero}, \bibinfo{person}{Pietro Li{\`{o}}}, {and} \bibinfo{person}{Yoshua
  Bengio}.} \bibinfo{year}{2018}\natexlab{}.
\newblock \showarticletitle{Graph Attention Networks}. In
  \bibinfo{booktitle}{\emph{6th International Conference on Learning
  Representations, {ICLR} 2018, Vancouver, BC, Canada, April 30 - May 3, 2018,
  Conference Track Proceedings}}. \bibinfo{publisher}{OpenReview.net}.
\newblock


\bibitem[Wan et~al\mbox{.}(2022)]%
        {wan2022they}
\bibfield{author}{\bibinfo{person}{Yao Wan}, \bibinfo{person}{Wei Zhao},
  \bibinfo{person}{Hongyu Zhang}, \bibinfo{person}{Yulei Sui},
  \bibinfo{person}{Guandong Xu}, {and} \bibinfo{person}{Hai Jin}.}
  \bibinfo{year}{2022}\natexlab{}.
\newblock \showarticletitle{What Do They Capture? - {A} Structural Analysis of
  Pre-Trained Language Models for Source Code}. In
  \bibinfo{booktitle}{\emph{44th {IEEE/ACM} 44th International Conference on
  Software Engineering, {ICSE} 2022, Pittsburgh, PA, USA, May 25-27, 2022}}.
  \bibinfo{publisher}{{ACM}}, \bibinfo{pages}{2377--2388}.
\newblock


\bibitem[Willard and Louf(2023)]%
        {willard_efficient_2023}
\bibfield{author}{\bibinfo{person}{Brandon~T. Willard} {and}
  \bibinfo{person}{R{\'{e}}mi Louf}.} \bibinfo{year}{2023}\natexlab{}.
\newblock \showarticletitle{Efficient Guided Generation for Large Language
  Models}.
\newblock \bibinfo{journal}{\emph{CoRR}}  \bibinfo{volume}{abs/2307.09702}
  (\bibinfo{year}{2023}).
\newblock


\bibitem[Williams and Kessler(2003)]%
        {pairprogramming2}
\bibfield{author}{\bibinfo{person}{Laurie~A. Williams} {and}
  \bibinfo{person}{Robert~R. Kessler}.} \bibinfo{year}{2003}\natexlab{}.
\newblock \bibinfo{booktitle}{\emph{Pair Programming Illuminated}}.
\newblock \bibinfo{publisher}{Addison Wesley}.
\newblock
\showISBNx{978-0-201-74576-4}


\bibitem[Williams et~al\mbox{.}(2000)]%
        {pairprogramming1}
\bibfield{author}{\bibinfo{person}{Laurie~A. Williams},
  \bibinfo{person}{Robert~R. Kessler}, \bibinfo{person}{Ward Cunningham}, {and}
  \bibinfo{person}{Ron Jeffries}.} \bibinfo{year}{2000}\natexlab{}.
\newblock \showarticletitle{Strengthening the Case for Pair Programming}.
\newblock \bibinfo{journal}{\emph{{IEEE} Softw.}} \bibinfo{volume}{17},
  \bibinfo{number}{4} (\bibinfo{year}{2000}), \bibinfo{pages}{19--25}.
\newblock


\bibitem[Yun et~al\mbox{.}(2019)]%
        {graphtransformers}
\bibfield{author}{\bibinfo{person}{Seongjun Yun}, \bibinfo{person}{Minbyul
  Jeong}, \bibinfo{person}{Raehyun Kim}, \bibinfo{person}{Jaewoo Kang}, {and}
  \bibinfo{person}{Hyunwoo~J. Kim}.} \bibinfo{year}{2019}\natexlab{}.
\newblock \showarticletitle{Graph Transformer Networks}. In
  \bibinfo{booktitle}{\emph{Advances in Neural Information Processing Systems
  32: Annual Conference on Neural Information Processing Systems 2019, NeurIPS
  2019, December 8-14, 2019, Vancouver, BC, Canada}},
  \bibfield{editor}{\bibinfo{person}{Hanna~M. Wallach}, \bibinfo{person}{Hugo
  Larochelle}, \bibinfo{person}{Alina Beygelzimer}, \bibinfo{person}{Florence
  d'Alch{\'{e}}{-}Buc}, \bibinfo{person}{Emily~B. Fox}, {and}
  \bibinfo{person}{Roman Garnett}} (Eds.). \bibinfo{pages}{11960--11970}.
\newblock


\bibitem[Zhang et~al\mbox{.}(2023)]%
        {zhang_instruction_2023}
\bibfield{author}{\bibinfo{person}{Shengyu Zhang}, \bibinfo{person}{Linfeng
  Dong}, \bibinfo{person}{Xiaoya Li}, \bibinfo{person}{Sen Zhang},
  \bibinfo{person}{Xiaofei Sun}, \bibinfo{person}{Shuhe Wang},
  \bibinfo{person}{Jiwei Li}, \bibinfo{person}{Runyi Hu},
  \bibinfo{person}{Tianwei Zhang}, \bibinfo{person}{Fei Wu}, {and}
  \bibinfo{person}{Guoyin Wang}.} \bibinfo{year}{2023}\natexlab{}.
\newblock \showarticletitle{Instruction Tuning for Large Language Models: {A}
  Survey}.
\newblock \bibinfo{journal}{\emph{CoRR}}  \bibinfo{volume}{abs/2308.10792}
  (\bibinfo{year}{2023}).
\newblock


\bibitem[Zhou et~al\mbox{.}(2020)]%
        {zhou_graph_2020}
\bibfield{author}{\bibinfo{person}{Jie Zhou}, \bibinfo{person}{Ganqu Cui},
  \bibinfo{person}{Shengding Hu}, \bibinfo{person}{Zhengyan Zhang},
  \bibinfo{person}{Cheng Yang}, \bibinfo{person}{Zhiyuan Liu},
  \bibinfo{person}{Lifeng Wang}, \bibinfo{person}{Changcheng Li}, {and}
  \bibinfo{person}{Maosong Sun}.} \bibinfo{year}{2020}\natexlab{}.
\newblock \showarticletitle{Graph neural networks: {A} review of methods and
  applications}.
\newblock \bibinfo{journal}{\emph{{AI} Open}}  \bibinfo{volume}{1}
  (\bibinfo{year}{2020}), \bibinfo{pages}{57--81}.
\newblock


\end{thebibliography}

\end{document}